\documentclass[runningheads,a4paper]{llncs}
\usepackage[misc]{ifsym}
\usepackage{amssymb}
\usepackage{graphicx}
\usepackage[colorlinks,urlcolor=blue,citecolor=blue]{hyperref}
\usepackage{CJK}
\usepackage{booktabs}
\usepackage{url}
\newcommand{\keywords}[1]{\par\addvspace\baselineskip
\noindent\keywordname\enspace\ignorespaces#1}
\usepackage{bbding}
\usepackage{amsmath}

\begin{document}

\mainmatter  

\title{The Twin Conjugacy Search Problem and Applications}

\titlerunning{The Twin Conjugacy Search Problem and Applications}

%
%
\author{Xiaoming Chen$^1$, Weiqing You$^2$%
}
\authorrunning{X. Chen et al}

\institute{$^{1,2}$ Beijing Electronic Science \& Technology Institute
Beijing 100070, China\\
$^1$ University of Science and Technology of China, Hefei 230026, China \\
chenxmphd@yeah.net, 894572560@qq.com }

%
%

\toctitle{Lecture Notes in Computer Science}
\tocauthor{Authors' Instructions}
\maketitle

\begin{abstract}
We propose a new computational problem over the noncommutative group, called the twin conjugacy search problem. This problem is related to the conjugacy search problem and can be used for almost all of the same cryptographic constructions that are based on the conjugacy search problem. However, our new problem is at least hard as the conjugacy search problem.
Moreover, the twin conjugacy search problem have many applications. One of the most important applications, we propose a trapdoor test which can replace the function of the decision oracle. We also show other applications of the problem, including: a non-interactive key exchange protocol and a key exchange protocol, a new encryption scheme which is secure against chosen ciphertext attack, with a very simple and tight security proof and short ciphertexts, under a weak assumption, in the random oracle model.


\end{abstract}

\section{Introduction}
\subsection{Background and related work}

The conjugacy search problem is a more important computational problem on noncommutative groups while it is also a widely used cryptographic primitive. In the context of quantum computing, Ko\cite{jour1} first proposed a public key encryption system based on the design of conjugate problems on the braid group, which greatly promoted the development of the group theoretic cryptography. It makes more and more scholars focus on the research of the conjugacy search problem and the public key cryptosystem and get many excellent results\cite{jour2,jour3,jour4,jour5,jour200}.
Although the security of many cryptographic schemes on the braid group has been questioned\cite{jour6,jour7,jour12}, non commutative groups have attracted increasing attention\cite{jour8,jour9,jour300}.

The rapid development of quantum computing technology has aggravated the threat to the existing public key cryptosystem\cite{jour10,jour11}. However, the algorithm of resisting quantum attack is also constantly proposed. It may be an effective method to find the anti quantum attack algorithm on the non commutative group. As one of the difficult problems on the noncommutative group, the conjugacy search problem is very suitable for designing public key cryptosystems. In recent years, many good papers have been put forward\cite{jour13,jour14,jour15}.

Recently, inspired by David Cash et al. \cite{jour16}, we found that the conjugacy search problem has very similar results with the Diffie-Hellman problem, due to space constraints we must defer the details of the theory.

\subsection{The Diffie-Hellman problem and the Conjugacy Search problem}

To illustrate the similarities between Diffie-Hellman problem and conjugacy search problem more intuitively, show that:

\noindent \textbf{Hashed-ElGamal Encryption Scheme Based on Diffie-Hellman Problem} \\

\begin{tabular}{|p{11cm}| }
  \hline
  \\
\quad \underline{$KeyGen^{DH-hElG}(K):$}                                       \\
\quad $(x,g,X) \leftarrow GenKey_{DH}(K) $; \\
\quad $pk=(g,X), sk = x$, where $ X = g^x $. \\

\\

\quad \underline{$E_{pk}(m):$} \\
\quad $ y \leftarrow _R Z_q, Y = g^y, Z = X^y, k = H(Y,Z), c = E_k(m); $ \\
\quad output $(Y,c)$.  \\

\\

\quad \underline{$ D_{sk}(Y,c): $} \\
\quad $ Z = Y^x, k = H(Y,Z), m = D_k(c)$; \\
\quad output $m$. \\
\\
  \hline
\end{tabular} \\  \\

\noindent \textbf{Hashed-ElGamal Encryption Scheme Based on Conjugacy Search Problem}\cite{jour100} \\

\begin{tabular}{|p{11cm}| }
  \hline
  \\
\quad \underline{$KeyGen^{CSP-hElG}(K):$}                                       \\
\quad $(x,g,X) \leftarrow GenKey_{CSP}(K) $; \\
\quad $pk=(g,X), sk = x$, where $ X = xgx^{-1} $. \\

\\

\quad \underline{$E_{pk}(m):$} \\
\quad $ y \leftarrow _R G, Y = ygy^{-1}, Z = yXy^{-1}, k = H(Y,Z), c = E_k(m); $ \\
\quad output $(Y,c)$.  \\

\\

\quad \underline{$ D_{sk}(Y,c): $} \\
\quad $ Z = xYx^{-1}, k = H(Y,Z), m = D_k(c)$; \\
\quad output $m$. \\
\\
  \hline
\end{tabular} \\  \\

The ElGamal scheme\cite{jour17} based on the Diffie-Hellman problem is a great discovery in public key cryptography. The research in this aspect is more mature than the ElGamal scheme based on the conjugacy search problem. For the sake of conciseness, simplify the Hashed-ElGamal Encryption Scheme\cite{jour18} to DH-hElG. With the appropriate modification of the algorithm proposed by Ko\cite{jour1}, we get the Hashed-ElGamal Encryption Scheme based on the conjugacy search problem and simplify it to CSP-hElGamal.

Formally, there are many similarities between DH-hElG and CSP-hElG, and the main difference lies in the computational characteristics of groups. David propose a new computational problem called the twin Diffie-Hellman problem, and its applications\cite{jour16}, it has solved important problems on the security proofs of the Diffie-Hellman problem. Their results are very useful and amazing. Inspired by the work of David Cash et al.\cite{jour16}, we find some new properties of the conjugacy search problem on the noncommutative group.

\subsection{Our result}


In order to describe our work more succinctly, the braid group is used as the implementation group of this theory, but all of our results can be applied to any non commutative group, as long as the conjugacy search problem over it is difficult and it has two exchangeable subgroups. In this paper, for the first time, we define several security assumptions related to the conjugacy search problem, and analyze the security of the CSP-hElG scheme under each security assumptions. the main results are as follow:

\begin{enumerate}
  \item We propose a new computation problem called the twin conjugacy search problem, and prove that it is at least as hard as the ordinary conjugacy search problem. When give a decision oracle, our new problem is still hard.
  \item We show that a trapdoor test based on the twin conjugacy search problem, and it can simulate the function of a decision oracle.
  \item A non-interactive key exchange protocol and a key exchange protocol based on the new problem are proposed, and we show that a new public key encryption scheme which is secure against chosen ciphertext attack. Our proof just need to a weak assumption, in the random oracle medel.
\end{enumerate}

\section{Preliminaries}

Braid group is a typical non-commutative group, it is an important way that using braid group as an implementation tool to explore cryptography algorithm on non commutative group. There are many studies on the braid group and the theory of cryptography\cite{jour19,jour20}. However, this paper only takes it as a implementation tool, and we no longer spend too much space on the braid group. At the same time, the reader does not have to fall into complex group theory. What we have to explain is:

Difine $B_n$ as a braid group generated by $\sigma_1, \sigma_2, \cdots, \sigma_{n-1}$, and if

\begin{displaymath}
    \left\{ \begin{array}{cc}
    \sigma_i\sigma_j\sigma_i=\sigma_j\sigma_i\sigma_j & \textrm{    \indent if $|i-j|=1$}\\
    \sigma_i\sigma_j=\sigma_j\sigma_i & \textrm{   \indent if $|i-j|>2$} \\
    \end{array} \right.
\end{displaymath}

Let $LB_l$ be the left subgroup of $B_n$, and let $RB_r$ be the right subgroup of $B_n$, the following algebraic operations need to be noted

$$ \forall x \in LB_l, \forall y \in RB_r, then \quad xy = yx $$

Conjugacy search problem is a very important computational problem in non-commutative group. It is to be known that the problem is hard on the braid group\cite{jour8,jour9}. Describe the following

\subsubsection{Conjugacy Search Problem (CSP)} $(g,X) \in B_n \times B_n,$, find $x \in B_n$ such that $X=xgx^{-1}$

Compared with other groups, braid group has richer connotations and algebraic properties. However, the braid group is not necessary for our theory, and it just for the conciseness of the narrative. In fact, our theory can be extended to any noncommutative group as long as the conjugacy search problem is difficult over it, and it have two commutative subgroups.

\section{Security Assumptions}

The security of public key cryptography rely on the security assumptions related to a difficult problem. Here are some security assumptions associated with the conjugate problem.

\subsubsection{The Computional Conjugacy Search Assumption(The CCS Assumption):} We assume that it is hard to compute $Z$, given the values $X$ and $Y$ in Braid group $B_{l+r}$. Define $z:=ccs(X,Y)$, where $X=xgx^{-1}, Y=ygy^{-1}, Z=xyg(xy)^{-1}$

Ko et al proposed a new public-key cryptosystem based on the braid group, we make some improvements to it, and name it as the conjugacy search encryption scheme(the CS encryption schme), that is, the CSP-hElG Scheme in the Introduction.

\subsubsection{The CS encryption scheme:} (Enc,Dec) is a pair of symmetric key encryption algorithms while $H$ is a hash function,$H:B_{l+r}\rightarrow \{0,1\}^{l(k)}$, $l(k)$ is a security parameter. $g$ is an element in $B_{l+r}$.

\begin{enumerate}
  \item \textbf{KeyGeneration} One chooses a random element $x$ in $LB_l$, compute $X=xgx^{-1}$, then the public key is $(X, g)$, while the private key is $(x,g)$.
  \item \textbf{Encryption} For cipher message $m\in B_{l+r}$, one chooses a random element $y$ in $RB_r$, compute
 $Y=ygy^{-1}, Z=yXy^{-1}, k=H(Y,Z), c=Enc_k(m)$. The ciphertext is $(Y,c)$.
  \item \textbf{Decryption} Decipher get the target ciphertext $(Y,c)$, compute $Z=xYx^{-1}, k=H(Y,Z),m=Dec_k(c)$.
\end{enumerate}

It has been proved that the CS encryption scheme is secure against chosen plaintext attack. However, the CCS assumption is not sufficient to establish the security of chosen ciphertext attack, even the $H$ is a random oracle. To illustrate the problem, an adversary selects group elements $\hat{Y},\hat{Z}$ randomly, to encrypt a message $m$, compute $\hat{k}=H(\hat{Y},\hat{Z})$, and $\hat{c}=Enc_{\hat{k}}(\hat{m})$. Futher, assume that the adversary gives the ciphertext $\hat{Y},\hat{c}$ to a decryption oracle obtaining the decryption $m$. It is easy to judge the equation
$\hat{Z} \overset{?}{=} H(X,\hat{Y})$ through the equation $m \overset{?}{=} \hat{m}$. So, for random elements $\hat{Y},\hat{Z}$, the adversary can answer $\hat{Z} \overset{?}{=} ccs(X,\hat{Y})$ through the decryption oracle, and the adversary can not do it on his own. So the adversary can break the scheme easily by forge a legal ciphertext.

In fact, when the adversary get a decryption oracle, what he need to do is compute $ccs(X,Y)$, after answering questions of the form $'is \ \hat{Z} \overset{?}{=} ccs(X,\hat{Y})'$ many times. Thus, we need a stronger assumption to ensure the security of Chosen-Ciphertext Attack(CCA).

\subsubsection{The Strong CCS Assumption:} We assume that it is hard to compute $css(X,Y)$, given random $X, Y$ in $B_{l+r}$, along with access to a decision oracle for the predicate $ccsp(X, \cdot , \cdot)$, which on input $(\hat{Y},\hat{Z})$, returns $ccsp(X, \hat{Y},\hat{Z})$, define the predicate

$$ccsp(X, \hat{Y},\hat{Z}):=ccs(X,\hat{Y}) \overset {?}{=} \hat{Z}$$

It is not difficult to prove that the CS encryption scheme is secure against chosen ciphertext attack when the $H$ is modeled as a random oracle, under the strong CCS assumption and if the underlysing symmetric cipher $(Enc,Dec)$ is itself secure against chosen ciphertext attack.$cite my paper$

Compare to the CCS assumption, The Strong CCS assumption is too stronger. In genral, the weaker the assumption, the more secure the algorithm is, and the results are more rigorous. To get CCA security under the CCS assumption, we propose a new computational problem:

\subsubsection{The Twin Conjugacy Search Problem(the twin CSP):} $ (g,X_1,X_2) \in B_n \times B_n \times B_n $, find $x_1,x_2 \in B_n$ such
that $X_1=x_1gx_1^{-1}, X_1=x_1gx_1^{-1}$.

Like the CS encryption scheme, we propose a new encryption scheme based on the twin conjugacy search problem.

\subsubsection{The Twin CS encryption scheme:} (Enc,Dec) is a pair of symmetric key encryption algorithms while $H$ is a hash function,$H:B_{l+r}\rightarrow \{0,1\}^{l(k)}$, $l(k)$ is a security parameter. $g$ is an element in $B_{l+r}$.

\begin{enumerate}
  \item \textbf{KeyGeneration} One chooses random elements $x_1,x_2$ in $LB_l$, compute $X_1=x_1gx_1^{-1},X_2=x_2gx_2^{-1}$, then the public key is $(X_1,X_2,g)$, while the private key is $(x_1,x_2)$.
  \item \textbf{Encryption} For cipher message $m\in B_{l+r}$, one chooses a random element $y$ in $RB_r$, compute
 $Y=ygy^{-1}, Z_1=yX_1y^{-1},Z_2=y X_2 y^{-1}, k=H(Y,Z), c=Enc_k(m)$. The ciphertext is $(Y,c)$.
  \item \textbf{Decryption} Decipher get the target ciphertext $(Y,c)$, compute $Z_1=x_1Yx_1^{-1},Z_2=x_2Yx_2^{-1}, k=H(Y,Z),m=Dec_k(c)$.
\end{enumerate}

Like the conjugacy search problem, we present the security assumption related the twin conjugacy search problem.

\subsubsection{The Twin Computational Conjugacy Search Assumption (The twin CCS assumption):} Suppose that it is hard to compute $Z_1,Z_2$ in braid group $B_{l+r}$, given the values $X_1,X_2,Y$ in braid group $B_{l+r}$. Define
 $$ (Z_1,Z_2):=2ccs(X_1,X_2,Y)=(ccs(X_1,Y),ccs(X_2,Y)) $$
where $X_1=x_1gx_1^{-1},X_2=x_2gx_2^{-1},Y=ygy^{-1},Z_1=(x_1y)g(x_1y)^{-1},Z_2=(x_2y)g(x_2y)^{-1}$ \\

In addition, we can present a stronger assumpution.
\subsubsection{The Strong Twin Computational Conjugacy Search Assumption (The Strong Twin CCS assumption):}
Suppose that it is hard to compute
$2ccs(X_1, \\ X_2,Y)$, given the values $X_1,X_2,Y$ in braid group $B_{l+r}$, along with access to a decision oracle for the predicate
$ccsp(X_1,X_2,\cdot,\cdot,\cdot)$, which on input $\hat{Y},\hat{Z_1},\hat{Z_2}$, returns $2ccsp(X_1,X_2,\hat{Y},\hat{Z_1},\hat{Z_2})$.
Define the predicate
 $$ 2ccsp(X_1,X_2,\hat{Y},\hat{Z_1},\hat{Z_2}) := ( ccs(X_1, Y), ccs(X_2, Y) ) = (Z_1, Z_2) $$

It's easy to know that the twin CS encryption scheme is secure against chosen ciphertext attack when the $H$ is modeled as a random oracle, under the strong twin CCS assumption and if the underlysing symmetric cipher $(Enc,Dec)$ is itself secure against chosen ciphertext attack.

In above, we propose two encryption schemes and four kinds of security assumptions related to the conjugate problem over the braid group. Next we'll discuss the relationships of each assumption, and the security of the twin CS encryption scheme. One of our main results is the following:

\subsubsection{Theorem 1.} The CCS assumption holds if and only The strong twin CCS assumption holds.

It is not hard to see that the twin strong CCS assumption implies the CCS assumption, while the non-trivial direction to prove is that the CCS assumption implies the strong twin CCS assumption. However, we need to defer the proof of Theorem $1$ to save space.

\section{A Trapdoor Test and a Proof of Theorem 1}

In the following, we will propose another one of our results: Trapdoor test theorem. Our theory is largely inspired by David\cite{jour16}. David et al proposed the twin Diffie-Hellman problem and a trapdoor test over general cyclic group, their works are amazing, and it has greatly promoted the development of provable security theory. However, our theory is based on the conjugacy search problem over the non-commutative group. We emphasize

\subsubsection{Lemma} $\forall \ x,y \in B_n$, Remember $xy^{-1}$ as $\frac{x}{y}$,
$\forall r \in B_n$, then $$ r( \frac{x}{y} )r^{-1} = \frac{rxr^{-1}}{ryr^{-1}} $$
\subsubsection{proof}:
$$\frac{rxr^{-1}}{ryr^{-1}} = rxr^{-1} \cdot (ryr^{-1}) = r(xy^{-1})r^{-1}=r\frac{x}{y}r^{-1} $$

Now, we propose the main theory:

\subsubsection{Theorem 2 (Trapdoor Test). }

Let $B_{l+r}$ be a braid group, $RB_r$ and $LB_l$ are right subgroup and left subgroup of the $B_{l+r}$. $g$ is a random elemet on the group of $B_{l+r}$, choice $X_1 \leftarrow _R B_{l+r}, \ r \leftarrow _R LB_l, \ s \leftarrow RB_r$, define a random variable $X_2 = \frac{sgs^{-1}}{rX_1r^{-1}}$, if $\hat{Y},\hat{Z_1},\hat{Z_2}$ are random elements on the $B_{l+r}$. Then we have:

$(a.)$ $X_2$ is uniformly distributed over $G$;

$(b.)$ $X_1$ and $X_2$ are independent;

$(c.)$ if $X_1 = x_1gx_1^{-1}, X_2 = x_2gx_2^{-1}, x_1,x_2 \in LR_l$, then the probability that the truth value of

\begin{equation}
  \hat{Z_2} \cdot r\hat{Z_1}r^{-1} = s\hat{Y}s^{-1}
\end{equation}

\noindent does not agree with the truth value of

\begin{equation}
 \hat{Z_1}=x_1\hat{Y}x^{-1} \wedge \hat{Z_2}=x_2\hat{Y}x_2^{-1}
\end{equation}

\noindent is negligible; moreover, if $(3)$ holds, then $(2)$ certainly holds.

\subsubsection{proof:} Observe that
 $$X_2 = \frac{sgs^{-1}}{rX_1r^{-1}}$$

The elements $s$ and $r$ are randomly selected from $RB_r$ and $LB_l$, respectively, and $X_1 \in B_{l+r}$. It is easy to verify that $X_2$ is uniformly distributed over $B_{l+r}$, and that $X_1, X_2, r$ are mutually independent, from which $(a.)$ and $(b.)$ follow. To prove $(c.)$, condition on fixed values of $X_1, X_2$, suppose that $\hat{Y}=ygy^{-1}, y \in RB_{r}$. If $(2)$ holds, because of

 $$X_2 = \frac{sgs^{-1}}{rX_1r^{-1}}$$

then
$$ y X_2 y^{-1} = \frac{ysgs^{-1}y^{-1}}{yrX_1r^{-1}y^{-1}} = \frac{s\hat{Y}s^{-1}}{ryX_1y^{-1}r^{-1}} $$

That is,
$$ s\hat{Y}s^{-1} = (yX_2y^{-1})(ryX_1y^{-1}r^{-1}) $$
So,
\begin{eqnarray*}
\hat{Z_2} \cdot r\hat{Z_1}r^{-1} & = & (x_2\hat{Y}x_2^{-1})r(x_1\hat{Y}x_1^{-1})r^{-1} \\
                 & = & y X_2 y^{-1}ry X_1 y^{-1}r^{-1} \\
                 & = & s \hat{Y} s^{-1}
\end{eqnarray*}
Thus, while $(2)$ holds, $(1)$ certainly holds. Conversely, if $(2)$ does not hold, we show that $(1)$ holds with a negligible probability. Observe that $(1)$
$$ s\hat{Y}s^{-1} = yX_{2}y^{-1} \cdot (ryX_1 y^{-1} r^{-1}) = \hat{Z_2} \cdot r\hat{Z_1}r^{-1} $$
So
\begin{equation}
\hat{Z_2}^{-1}\cdot{yX_2 y^{-1}} = \frac{r\hat{Z_1}r^{-1}}{ryX_1 y^{-1}r^{-1}} =  r \frac{\hat{Z_1}}{yX_1y^{-1}} r^{-1}
\end{equation}

It is not hard to see, if $ \hat{Z_1} = x_1\hat{Y}x_1^{-1}$ and $\hat{Z_2} \neq x_2\hat{Y}x_2^{-1} $, then $(3)$ certainly does not hold. This leaves us with the case $\hat{Z_1} \neq x_1\hat{Y}x_1^{-1}$. But in the case, the right hand side of $(3)$ is a random element of $B_{l+r}$ since $r$ is uniformly distributed over $LB_l$, but the left hand side is a fixed element of $B_{l+r}$. It is easy to see that the probability is negligible which select an element from $B_{l+r}$ to make it equal to a fixed element of $B_{l+r}$.

Now, we can prove the theorem $1$ through the trapdoor test.

\subsubsection{Theorem 1.} The CCS assumption holds if and only The strong twin CCS assumption holds.

\subsubsection{Proof:} The twin strong CCS assumption implies the CCS assumption obviously. To prove that the CCS assumption implies the strong twin CCS assumption. Let us define some terms:

assume that an adversary $B$ who attack the CCS assumption, an adversary $A$ who attack the strong twin CCS assumption. $B$ get the challenge instance $(X,Y)$ of the CCS assumption, the target is to compute $ccs(X,Y)$.

First, $B$ chooses $r \leftarrow _R LB_l, \ s \leftarrow _R RB_r$, sets
$$X_1=X, \ X_2 = \frac{sgs^{-1}}{rX_1r^{-1}} $$
and give $A$ the challenge instance $(X_1, X_2, Y)$, $A$ need to do to compute $(Z_1,Z_2)=2ccs(X_1,X_2,Y)$.

Second, $A$ chooses $\hat{Y},\hat{Z_1},\hat{Z_2}$ to query $B$, then $B$ processes each decision query $\hat{Y},\hat{Z_1},\hat{Z_2}$ by testing if
$\hat{Z_2}\cdot r\hat{Z_1}r^{-1} = s\hat{Y}s^{-1}$ holds.

Finally, if and when $A$ outputs $(Z_1,Z_2)$, $B$ tests if this output is correct by testing if $Z_2 \cdot rZ_1r^{-} = sYs^{-1}$ holds. If this does not hold, then $B$ outputs "failure", otherwise, $B$ outputs $Z_1$. The proof is easily completed using the trapdoor test.

\section{Twin CCS-ElGamal Cryptosystem}

\subsection{Security Model}
The security model is portrayed by Indistinguishability-Game (IND-GAME), mainly divided into three levels: Indistinguishability-Chosen Plaintext Attack (IND-CPA) \cite{jour21}, Indistinguishability - (Non Adaptive) Chosen Ciphertext Attack (IND-CCA) \cite{jour22}, Indistinguishability - (Adaptive) Chosen Ciphertext Attack (IND-CCA2) \cite{jour23}. We recall the definition for the CCA2.

\subsubsection{Definition  Indistinguishability - (Adaptive) Chosen Ciphertext Attack (IND-CCA2)}\cite{jour23}
The IND game of public key encryption scheme under (Adaptive) chosen ciphertext attack (IND-CCA2) is as follows \\
\begin{enumerate}
  \item Initialization. The Challenger $B$ generates the password system, and the Adversary $A$ obtains the system public key $pk$.
  \item Training1. $A$ sends the ciphertext $C$ to the $B$, and $B$ sends the decrypted plaintext to $A$.(Polynomial bounded)
  \item Challenge. The Adversary $A$ outputs two messages of the same length, $M_0$ and $M_1$. The Challenger $B$ chooses $\beta\leftarrow_R\{0,1\}$, cipher $M_{\beta}$, and send ciphertext $C^\ast$ (Target ciphertext) to $A$.
  \item Training2. $A$ sends the ciphertext $C(C\neq C^\ast)$ to the $B$, and $B$ sends the decrypted plaintext to $A$.(Polynomial bounded)
  \item Guess. $A$ output $\beta'$, if $\beta'=\beta$, return 1, $A$ attack success.
\end{enumerate}

The advantage of the adversary $A$ can be defined as a function of the parameter $K$:
$$ Adv_A^{CCA2}(K)=\left| Pr[\beta'=\beta]-\frac{1}{2} \right| $$
If exist a polynomial time adversary $A$, there is a negligible function $\varepsilon(K)$ that makes $Adv_A^{CCA2}(K)\leq \varepsilon(K)$ set up, it is called IND-CCA2 security.

\subsection{Key Exchange Protocol}
In the following we propose a new non-interactive key exchanege protocol based on the twin conjugacy search problem.

\subsubsection{Non-interactive Key Exchange Protocol:} Suppose that the two parties need to communicate are Alice and Bob. $g$ is a random element in braid group $B_{l+r}$. Alice's secret key is $(x_1,x_2)$, $x_1,x_2 \in LB_l$
pulic key is $(X_1,X_2)$, where $X_1=x_1 g x_1 ^{-1}, X_2=x_2 g x_2 ^{-1}$; Bob's secret key is $(y_1,y_2)$, $y_1,y_2 \in RB_r $, public key is $(Y_1,Y_2)$, where $Y_1=y_1 g y_1 ^{-1}, Y_2=y_2 g y_2 ^{-1}$. Keys which belong to Alice and Bob are authenticated by a trusted third party, they can share the key:
\begin{itemize}
  \item Alice compute \quad $x_1Y_1x_1^{-1},x_1Y_2x_1^{-1},x_2Y_1x_2^{-1},x_2Y_2x_2^{-1}$
  \item Bob compute   \quad $y_1X_1y_1^{-1},y_1X_2y_1^{-1},y_2X_1y_2^{-1},y_2X_2y_2^{-1}$
\end{itemize}

Because of $ccs(X_i,Y_j) = x_iY_jx_i^{-1} = y_jX_iy_j^{-1}, \ i = 1,2; \ j = 1,2 $, Alice and Bob can compute the same value through the same hash function $H$:
$$ k = H(ccs(X_1,Y_1),ccs(X_1,Y_2),ccs(X_2,Y_1),ccs(X_2,Y_2)) $$

\subsubsection{Key Exchange Protocol:} Suppose that the two parties need to communicate are Alice and Bob. $g$ is a random element in braid group $B_{l+r}$.

\begin{enumerate}
  \item Alice chooses random secret elements $x_1, x_2 \in LB_l$ and sends $(X_1, X_2)$ to Bob, where $X_1=x_1 g x_1 ^{-1},
   X_2=x_2 g x_2 ^{-1}$;
  \item Bob chooses random secret elements $y_1, y_2 \in LB_l$ and sends $(Y_1, Y_2)$ to Bob, where $Y_1=y_1 g y_1 ^{-1},
  Y_2=y_2 g y_2 ^{-1}$;
  \item Alice receives $ X_1, X_2 $ and compute \quad  $x_1Y_1x_1^{-1},x_1Y_2x_1^{-1},x_2Y_1x_2^{-1},x_2Y_2x_2^{-1}$;
  \item Bob receives $ Y_1, Y_2 $ and compute  \quad $y_1X_1y_1^{-1},y_1X_2y_1^{-1},y_2X_1y_2^{-1},y_2X_2y_2^{-1}.$
\end{enumerate}

Because of $ccs(X_i,Y_j) = x_iY_jx_i^{-1} = y_jX_iy_j^{-1}, \ i = 1,2; \ j = 1,2 $, Alice and Bob can compute the same value through the same hash function $H$:
$$ k = H(ccs(X_1,Y_1),ccs(X_1,Y_2),ccs(X_2,Y_1),ccs(X_2,Y_2)) $$

\subsection{The twin CS encryption scheme}

\subsubsection{Theorem 3.} Suppose that $H$ is modeled as a random oracle, The twin CS encryption scheme is secure against Chosen Ciphertext Attack under the CCS assumption and that the underlying symmetric cipher is itself secure against chosen ciphertext attack.

\subsubsection{Proof:} It is easy to see that the twin CS encryption scheme is secure against chosen ciphertext attack under the strong twin CCS assumption and that the underlying symmetric cipher is itself secure against chosen ciphertext attack, $H$ is modeled as a random oracle. However, according to theorem $1$, the CCS assumption holds if and only if the strong twin CCS assumption holds. So, The twin CS encryption scheme is secure against Chosen Ciphertext Attack under the condition of the theorem $3$.

\section{Conclusion}
Our work would like to avoid making Stronger assumptions, or working with specialized groups. All of the theory in this paper built in the braid group, however, our theory applies to any noncommutative group as long as the conjugacy search problem is hard over it. Compare to the original CSP-scheme, we make a little changes in the process of the encryption. In fact, the main achievement of this paper is to extend the conclusion which the twin Diffie-Hellman problems on general cyclic groups proposed by David Cash to the twin Conjugate Search Problems on general noncommutative groups.

\subsubsection*{Acknowledgments.} We thank any reviewers to comments our paper.

\end{document}